\documentclass[%
 onecolumn,
superscriptaddress,
nofootinbib,
 amsmath,amssymb,
 aps,d,10pt
]{revtex4-2}
\usepackage{graphicx}
\usepackage{float}
\usepackage{siunitx}
\usepackage{setspace}
\usepackage{dcolumn}
\usepackage{bm}
\usepackage{xcolor}
\usepackage{url}
\usepackage{tikz}

\usepackage{lineno}

\renewcommand{\bell}{\boldsymbol{\ell}}
\usepackage{float}
\renewcommand{\vec}[1]{\mathbf{#1}}

\def\fr#1{\textcolor{black}{[#1]}}

\onecolumngrid
\usepackage{comment}
\usepackage[colorlinks=true,linkcolor=magenta,citecolor=magenta]{hyperref}%

\begin{document}


\title{CMB lensing with shear-only reconstruction on the full sky\\}

\author{Frank J. Qu}
\email{jq247@cam.ac.uk}
\affiliation{DAMTP, Centre for Mathematical Sciences, Wilberforce Road, Cambridge CB3 0WA, UK}
\affiliation{Kavli Institute for Cosmology Cambridge, Madingley Road, Cambridge, CB3 0HA, UK}

\author{Anthony Challinor}
\email{a.d.challinor@ast.cam.ac.uk}
 \affiliation{DAMTP, Centre for Mathematical Sciences, Wilberforce Road, Cambridge CB3 0WA, UK}
\affiliation{Institute of Astronomy, Madingley Road, Cambridge, CB3 0HA, UK}
\affiliation{Kavli Institute for Cosmology Cambridge, Madingley Road, Cambridge, CB3 0HA, UK}

\author{Blake D.\ Sherwin}
\email{sherwin@damtp.cam.ac.uk}
\affiliation{DAMTP, Centre for Mathematical Sciences, Wilberforce Road, Cambridge CB3 0WA, UK}
\affiliation{Kavli Institute for Cosmology Cambridge, Madingley Road, Cambridge, CB3 0HA, UK}

\date{\today}

\begin{abstract}
 Reconstruction of gravitational lensing effects in the CMB from current and upcoming surveys is still dominated by temperature anisotropies. Extragalactic foregrounds in temperature maps can induce significant biases in the lensing power spectrum obtained with the standard quadratic estimators.  Techniques such as masking cannot remove these foregrounds fully, and the residuals can still lead to large biases if unaccounted for. In this paper, we study the ``shear-only'' estimator, an example of a class of geometric methods that suppress extragalactic foreground contamination while making only minimal assumptions about foreground properties. The shear-only estimator has only been formulated in the flat-sky limit and so is not easily applied to wide surveys. Here, we derive the full-sky version of the shear-only estimator and its generalisation to an $m=2$ multipole estimator that has improved performance for lensing reconstruction on smaller scales. The multipole estimator is generally not separable, and so is expensive to compute. We explore separable approximations based on a singular-value decomposition, which allow efficient evaluation of the estimator with real-space methods. Finally, we apply these estimators to simulations that include extragalactic foregrounds and verify their efficacy in suppressing foreground biases.
\end{abstract}

\maketitle


\section{\label{sec:Introduction}Introduction}
    Gravitational lensing of the CMB encodes a wealth of information about our Universe. Observing the deflections produced by the intervening large-scale structure on the paths of CMB photons allows us to make integrated measurements of the projected matter distribution to high redshifts \cite{LEWIS_2006}.  CMB lensing provides us with a powerful probe to constrain parameters such as neutrino masses \cite{LESGOURGUES_2006,Qu2022} and dark energy \cite{Calabrese_2009}. Analyses with Planck and AdvACT data, building on earlier work by the Atacama Cosmology Telescope (ACT) and South Pole Telescope \cite{Smith:2007rg,Das_2011}, have demonstrated the great potential of this approach; see~\cite{Planck:2018lbu,Carron:2022eyg,qu2023atacama} for the most recent results. Current and future surveys, such as AdvACT \cite{Henderson_2016}, SPT-3G \cite{2014}, Simons Observatory \cite{Ade_2019} and CMB-S4 \cite{abazajian2016cmbs4}, will improve the precision of CMB lensing measurements significantly, making the identification and reduction of systematic biases increasingly important. While one expects polarisation information to dominate in the reconstruction of lensing from future surveys, many current and upcoming CMB surveys will still rely heavily on temperature. In this regime, extragalactic foreground contamination from the cosmic infrared background (CIB), the thermal Sunyaev--Zel'dovich effect (tSZ), the kinematic Sunyaev--Zel'dovich effect (kSZ) and radio point sources (PS) can leak into the lensing estimator producing significant biases if unaccounted for \cite{van_Engelen_2014}. Several mitigation methods for these biases have been proposed. For example, masking out sources from a known catalogue can decrease this bias, and techniques such as bias hardening \cite{Namikawa2013,Sailer2020}, which involves reconstructing and projecting out foregrounds, are useful for cases when the statistical properties of the foregrounds are known. Another method is multi-frequency component separation \cite{Madhavacheril_2018}, which can reduce or null specific foregrounds, but it was found that simultaneously reducing the CIB and tSZ increases the noise by a large factor. An improved technique, building upon \cite{Madhavacheril_2018}, was introduced in \cite{Darwish_2020} to eliminate foregrounds from the tSZ while preserving most of the signal-to-noise. Finally, \cite{darwish2021optimizing,sailer2021optimal} explore which combinations of multi-frequency cleaning and geometric methods (bias hardening and shear) are most effective in controlling lensing biases with only modest reduction in signal-to-noise.

In this paper, we focus on the shear estimator introduced in \cite{Schaan2019}, which built on earlier work exploring the role of magnification and shear in CMB lensing reconstruction~\cite{2018,PhysRevD.85.043016}. The idea behind the shear estimator is to exploit the different geometric effects on the local CMB two-point function of lensing and extragalactic foregrounds to separate them. In the limit where large-scale lenses are reconstructed from small-scale temperature anisotropies, weak lensing produces local distortions in the 2D CMB power spectrum with an isotropic part (i.e., monopole) due to lensing convergence and a quadrupolar part due to lensing shear.

The quadratic estimators usually employed in lensing reconstruction~\cite{Hu_2002,Okamoto2003} optimally combine convergence and shear in the large-scale-lens limit. However, extragalactic foreground power predominantly biases the local monopole power spectrum, leaving the shear-only estimator much less affected by foregrounds than the standard quadratic estimator. Moreover, with the shear-only estimator, one can include smaller-scale temperature modes in the reconstruction without introducing unacceptable levels of bias, thus mitigating the loss of signal-to-noise from discarding convergence information in the case of high-resolution, low-noise observations~\cite{Schaan2019}.


For the reconstruction of smaller-scale lenses, it is no longer true that the lensing convergence and shear can be considered constant over the coherence scale of the CMB. In this limit, lensing not only introduces monopole ($m=0$) and quadrupole ($m=2$) couplings in the local two-point function of the lensed CMB but also higher-order couplings. Furthermore, the dependencies of the $m=0$ and $m=2$ couplings on the angular scale of the CMB fluctuations and lenses deviate significantly from their large-lens limits. For reconstruction on smaller scales, one can formulate a set of multipole estimators, each extracting information from a specific $m$~\cite{Schaan2019}. Most of the reconstruction signal-to-noise is still contained in the $m=0$ and $m=2$ estimators, and extragalactic foregrounds are expected still to bias mainly $m=0$. However, it is necessary to use the correct scale dependence of the $m=2$ estimator to avoid the poor performance of the shear estimator when it is extended directly to reconstruction of smaller-scale lenses. This makes efficient evaluation with real-space methods difficult since the $m=2$ estimator is not generally separable.

    The shear reconstruction discussed in \cite{Schaan2019} is based on the flat-sky approximation. Since current and future high-resolution CMB experiments will cover a significant fraction of the sky, a full-sky formulation of the shear estimator is required. In this paper, we derive the full-sky version of the shear estimator and show how it can be evaluated efficiently in real space with spin-weighted spherical harmonic transforms. We generalise further to an $m=2$ multipole estimator to avoid the sub-optimal performance of the shear estimator on smaller scales. We suggest a simple separable approximation for this estimator using singular-value decomposition, which allows efficient evaluation with real-space methods. \fr{Although extending the shear formalism to polarization-based estimators is straightforward, as shown very recently in the flat-sky limit by \cite{PhysRevD.107.023504}, our focus in this paper is solely on the shear formalism for the temperature part of the lensing estimator.  This is motivated by the fact that the dominant sources of extragalactic foregrounds from the CIB or tSZ are 
    only weakly polarized compared to the CMB fluctuations on the scales that dominate lensing reconstruction, meaning that the biases they induce are relatively smaller in polarization-based reconstructions than for temperature. Moreover, temperature-based estimators still make a significant contribution to lensing reconstruction at the sensitivity of current surveys. While lensing with future surveys, such as CMB-S4~\cite{abazajian2016cmbs4}, will instead be dominated by polarization and will achieve much higher signal-to-noise, the main contributor -- the $EB$ quadratic estimator -- is effectively a shear estimator and therefore naturally suppresses foreground biases.}


This paper is organised as follows. In Sec. \ref{theory} we review CMB lensing reconstruction and multipole estimators in the flat-sky approximation and show how to generalise these to the curved sky. For the full-sky shear estimator, we
show that it provides an unbiased recovery of the lensing power spectrum using lensed CMB maps. Section~\ref{svd} then explores ways of improving the shear estimator by constructing a separable approximation to the $m=2$ estimator using singular-value decomposition. Finally, in Sec. \ref{test} we test the efficacy of the full-sky shear and $m=2$ estimators in suppressing extragalactic foreground contamination by measuring the lensing power spectrum using CMB simulations injected with foregrounds from the Websky simulation~\cite{Stein2020}.
in Appendix~\ref{appA} we discuss the estimator normalisation and reconstruction noise power.
In Appendix~\ref{appB} we derive the form of the full-sky shear estimator in spherical-harmonic space. 

\section{Theory: Shear and multipole estimators}\label{theory}

In this section, we introduce the standard lensing quadratic estimator and decompose it into a series of multipole estimators in the flat-sky limit. Starting from this, we then derive the full-sky form of the multipole ($m=2$) and shear estimators.

\subsection{Multipole and shear estimators in the flat-sky limit}

Lensing produces local distortions in the CMB two-point function, breaking the statistical isotropy of the unlensed CMB field and hence introducing new correlations between different Fourier modes over a range of wavenumbers defined by the lensing potential. Averaging over an ensemble of temperature fields for a fixed lensing potential $\phi$ results in off-diagonal correlations in the observed temperature field $T$:
\begin{equation}
    \langle T(\bell_1)T(\bell_2)\rangle_{\text{CMB}}=(2\pi)^2 \delta^{(2)}(\bell_1+\bell_2) C_{\ell_1}^{TT} + f^\phi(\bell_1,\bell_2) \phi(\bell_1 + \bell_2) \, ,
\end{equation}
with $f^{\phi}(\bell_1,\bell_2)=(\bell_1+\bell_2)\cdot \left(\bell_1 C^{TT}_{\ell_1}+\bell_2C^{TT}_{\ell_2}\right)$ and $C^{TT}_{\ell}$ the temperature power spectrum.\footnote{We use an improved version of the lensing response function $f^\phi(\bell_1,\bell_2)$ that describes the linear response of the CMB two-point function to variation in the lensing potential $\phi(\Vec{L})$, averaging over CMB and other lenses~\cite{Lewis:2011fk}. In particular, we use the lensed CMB power spectrum rather than the unlensed spectrum in $f^\phi$, which gives a good approximation to the true non-perturbative response function.}
The standard quadratic estimator~\cite{Hu_2002} exploits the coupling of otherwise-independent temperature modes to reconstruct the lensing field $\hat{\phi}$ by combining pairs of appropriately filtered temperature fields:
\begin{equation}
    \hat{\phi}^{\text{QE}}(\Vec{L})=\frac{A^{\text{QE}}_L}{2}\int\frac{d^2\bell}{(2\pi)^2}\frac{T(\bell+\Vec{L}/2)}{C^{\text{total}}_{|\bell+\Vec{L}/2|}} \frac{T(\Vec{L}/2-\bell)}{C^{\text{total}}_{|\Vec{L}/2-\bell|}} f^\phi(\bell+\Vec{L}/2,\Vec{L}/2-\bell) \, ,
\end{equation}
where $A^{\text{QE}}_L$ is a multipole-dependent normalisation to make the reconstructed field unbiased and  ${C}_{\bell}^{\text{total}}$ denotes the total temperature power spectrum, including residual foregrounds and instrumental noise. We follow the standard practice of using upper-case $L$ to denote a lensing multipole and lower-case $\ell$ to refer to CMB multipoles.

The angular dependence of the lensing response function can be expanded in a Fourier series in $\theta_{\Vec{L},\bell}$, the angle between $\Vec{L}$ and $\bell$:
\begin{equation}\label{mulres}
    f^\phi(\bell+\Vec{L}/2,\Vec{L}/2-\bell) = \sum_{\text{$m$ even}}f^m_{L,\ell}\cos(m\theta_{\vec{L},\bell})\, , \quad \text{with} \quad f^m_{L,\ell}= \begin{cases}
    \frac{1}{2\pi}\int d\theta_{\Vec{L},\bell} f^\phi(\bell+\vec{L}/,\vec{L}/2-\bell) & \text{if $m=0$,}\\
    \frac{1}{\pi}\int d\theta_{\Vec{L},\bell} f^\phi(\bell+\vec{L}/2,\vec{L}/2-\bell)\cos (m\theta_{\Vec{L},\bell})            & \text{otherwise.}
\end{cases}
\end{equation}
The expansion only involves even multipoles $m\in2\mathbb{N}$ because $f^\phi(\bell+\Vec{L}/2,\Vec{L}/2-\bell)$ is invariant under $\Vec{L} \rightarrow - \Vec{L}$, i.e., $\theta_{\Vec{L},\bell} \rightarrow \theta_{\Vec{L},\bell} + \pi$. In the limit that the lenses are on much larger scales than the CMB fluctuations they are lensing, $L \ll \ell$, the expansion~\eqref{mulres} is dominated by the $m=0$ (monopole) and $m=2$ (quadrupole) moments. The former corresponds to isotropic magnification or demagnification and the latter to shear. Expanding in $x \equiv L/\ell$, we have
\begin{equation}\label{response}
    f^\phi(\bell+\Vec{L}/2,\Vec{L}/2-\bell) =\frac{1}{2} L^2 C^{TT}_\ell \left[\left(\frac{d \ln \ell^2 C_\ell^{TT}}{d\ln \ell} + \frac{d \ln C_\ell^{TT}}{d\ln \ell} \cos 2 \theta_{\Vec{L},\bell}\right) + \mathcal{O}(x^2)\right] \, ,
\end{equation}
which involves only $m=0$ and $m=2$ terms at leading order.

The multipole expansion of the response function gives rise to a family of lensing estimators characterised by the multipole $m$, generally of the form
\begin{equation}
\hat{\phi}^{m}(\Vec{L}) = \frac{A_L^m}{2} \int \frac{d^2 \bell}{(2\pi)^2}\, g^m_{L,\ell} \cos (m \theta_{\Vec{L},\bell})
T(\bell+\Vec{L}/2) T(\Vec{L}/2-\bell) \, ,
\end{equation}
where the normalisation $A^m_L$ is chosen so that the estimator is unbiased:
\begin{equation}
1 = \frac{A^m_L}{2} \int \frac{\ell d \ell}{2\pi} \, g^m_{L,\ell} f^m_{L,\ell} \times \begin{cases} 1 & \text{if $m=0$} \\ 1/2 & \text{otherwise} \, .
\end{cases}
\label{eq:flatnorm}
\end{equation}
The multipole weight functions $g^m_{L,\ell}$ may be chosen in various ways. In Ref.~\cite{Schaan2019}, the minimum-variance estimator \emph{at each multipole} is constructed, in which case the $\ell$-dependence of $g^m_{L,\ell}$ follows
\begin{equation}\label{g_weight}
g^m_{L,\ell} \propto \left(\int \frac{d\theta_{\Vec{L},\bell}}{2\pi}\cos^2(m \theta_{\Vec{L},\bell}) C^{\text{total}}_{|\bell+\Vec{L}/2|}
C^{\text{total}}_{|\bell-\Vec{L}/2|}\right)^{-1} f^m_{L,\ell} \, ,
\end{equation}
where the integral here depends only on the magnitudes $\ell$ and $L$. We use a simpler form in this paper (with a relatively minor impact on optimality) whereby we replace $C^{\text{total}}_{|\bell\pm \Vec{L}/2|}$ appearing explicitly in the weight function in Eq.~\eqref{g_weight} with $C_\ell^{\text{total}}$, which is correct for their product to $\mathcal{O}(x^2)$. Reference~\cite{Schaan2019} show that most of the information in the lensing reconstruction is captured by the $m=0$ and $m=2$ multipole estimators, even for smaller-scale lenses where the squeezed limit $L \ll \ell$ does not apply.

It is convenient to split the QE estimator into this family of multipole estimators because some multipoles are more affected by foregrounds than others. The $m=2$ estimator, for instance, is expected to be more robust to extragalactic foregrounds since they primarily bias the $m=0$ estimator~\cite{Schaan2019}. We discuss this further in Sec.~\ref{subsec:foregrounds} below.

The above multipole estimators are generally not easy to implement efficiently because they are non-separable expressions of $L$ and $\ell$. To allow for fast evaluation with real-space methods, we first consider the squeezed-limit of the $m=2$ estimator. In this case, we use the approximate form of $f^2_{L,\ell}$ given by the leading-order quadrupole part of Eq.~\eqref{response}:
\begin{equation}
f^2_{L,\ell} \rightarrow f^{\text{shear}}_{L,\ell} = \frac{1}{2} L^2 C^{TT}_\ell  \frac{d \ln C_\ell^{TT}}{d\ln \ell} \, ,
\end{equation}
which is clearly separable in $L$ and $\ell$. We make a further simplication~\cite{Schaan2019}, replacing $T(\bell+\Vec{L}/2) T(\Vec{L}/2-\bell)$ with $T(\bell) T(\Vec{L}-\bell)$ to allow fast evaluation of the estimator. Note that this is not simply a variable transformation as we do not change the arguments of the weight function or the angle $\theta_{\Vec{L},\bell}$. The 
foreground deprojection argument still holds at leading order in this case (see Sec.~\ref{subsec:foregrounds}).
With these modifications, we obtain the shear estimator
\begin{equation}
\label{eq.flat}
      \hat{\phi}^{\text{shear}}(\vec{L})=\frac{A^{\text{shear}}_L}{2} \int\frac{d^2\bell}{(2\pi)^2} \, g^{\text{shear}}_{L,\ell} \cos (2\theta_{\Vec{L},\bell}) T(\bell)T(\vec{L}-\bell) \, ,
\end{equation}
where the shear weight function is
\begin{equation}\label{eq.flatw}
    g^{\text{shear}}_{L,\ell}=\frac{L^2}{2}\frac{C^{TT}_\ell}{(C^{\text{total}}_\ell)^2}\frac{d\ln{C^{TT}_\ell}}{d\ln{\ell}} \, .
\end{equation}
The shear normalisation is obtained from Eq.~\eqref{eq:flatnorm}. Equation~\eqref{eq.flat} can be evaluated efficiently by
first noting that the angular term $\cos (2 \theta_{\vec{L},\bell})$ can be expressed in terms of the contraction of the symmetric, trace-free tensors $\hat{L}_{\langle i} \hat{L}_{j\rangle}$ and $\hat{\ell}_{\langle i} \hat{\ell}_{j\rangle}$ (where overhats denote unit vectors in this context and the angular brackets denote the symmetric, trace-free part of the tensor):
\begin{align}
2\hat{L}_{\langle i} \hat{L}_{j\rangle} \hat{\ell}^{\langle i} \hat{\ell}^{j\rangle} & = 2\left(\hat{L}_{i} \hat{L}_{j} - \frac{1}{2} \delta_{ij} \right) \hat{\ell}^i \hat{\ell}^j \nonumber \\
&= 2\cos^2 \theta_{\Vec{L},\bell} - 1 \nonumber \\
&= \cos (2\theta_{\Vec{L},\bell}) \, .
\end{align}
We then simplify the shear estimator as follows:
\begin{align}
    \hat{\phi}^{\text{shear}}(\vec{L}) &= \frac{1}{2}{A_{L}^{\text{shear}}}L_{\langle i}L_{j \rangle}\int\frac{d^2\bell_1}{(2\pi)^2}\int\frac{d^2\bell_2}{(2\pi)^2}\, (2\pi)^2 \delta^{(2)}(\Vec{L}-\bell_1-\bell_2)
 \left(\frac{C^{{TT}}_{\ell_1}}{(C^{\text{total}}_{\ell_1})^2}\frac{1}{\ell_1^2}\frac{d\ln{C}^{{TT}}_{\ell_1}}{d\ln{\ell_1}}T(\bell_1)\right)\ell_1^{\langle{i}}\ell_1^{j\rangle}T(\bell_2) \nonumber \\
 &=   \frac{1}{2}{A_{L}^{\text{shear}}}L_{\langle i}L_{j \rangle}\int d^2\vec{x} \int\frac{d^2\bell_1}{(2\pi)^2}\int\frac{d^2\bell_2}{(2\pi)^2}e^{-i\Vec{L}\cdot\vec{x}}\left(\frac{C^{{TT}}_{\ell_1}}{(C^{\text{total}}_{\ell_1})^2}\frac{1}{\ell_1^2}\frac{d\ln{C}^{{TT}}_{\ell_1}}{d\ln{\ell_1}}T(\bell_1)\right)\ell_1^{\langle i}\ell_1^{j\rangle}e^{i\bell_1\cdot \vec{x}}T(\bell_2)e^{i\bell_2\cdot \vec{x}}\nonumber \\ &=-\frac{1}{2}A^{\text{shear}}_{L}L_{\langle i}L_{j \rangle}\int d^2\vec{x} e^{-i\vec{L}\cdot \vec{x}}T(\vec{x})\partial^{\langle i}\partial^{j\rangle} {T}^F(\vec{x}) \nonumber \\
 &= \frac{1}{2}A^{\text{shear}}_{L} \int d^2\vec{x} \left(\partial_{\langle i}\partial_{j \rangle}e^{-i\vec{L}\cdot \vec{x}}\right)T(\vec{x})\partial^{\langle i}\partial^{j\rangle} {T}^F(\vec{x}) \, ,\label{eq1111}
\end{align}
where the filtered temperature field is
\begin{equation}
{T}^{F}(\vec{x})=\int\frac{d^2\bell}{(2\pi)^2}\frac{C^{TT}_\ell}{(C^{\text{total}}_\ell)^2}\frac{1}{\ell^2}\frac{d\ln{C^{TT}_\ell}}{d\ln{\ell}}T(\bell) e^{i\bell\cdot \vec{x}} \, . 
\label{eq:flatshearfilter}
\end{equation}
The last line of Eq.~\eqref{eq1111} shows that the shear estimator is equivalent to extracting the $E$-mode part of the product of the temperature field and the symmetric, trace-free derivative of the filtered temperature field. Expressing the estimator in this form makes its translation to the curved sky rather straightforward, as we discuss in Sec.~\ref{sec:fullsky}.

The shear estimator can be evaluated very efficiently, and has excellent immunity to extragalactic foregrounds. However, approximating $f^2_{L,\ell}$ with its squeezed-limit $f^{\text{shear}}_{L,\ell}$ in the weight function results in poor performance at high $L$, where $L \ll \ell$ is not a good approximation~\cite{Schaan2019}. In Sec.~\ref{svd}, we suggest an alternative, separable approximation to $f^2_{L,\ell}$ that performs better than the shear estimator at high $L$. The approximation we develop there takes as its starting point the asymmetric form of the standard quadratic estimator:
\begin{equation}
    \hat{\phi}^{\text{QE}}(\Vec{L})=\frac{A^{\text{QE}}_L}{2}\int\frac{d^2\bell}{(2\pi)^2}\frac{T(\bell)}{C^{\text{total}}_{\ell}} \frac{T(\Vec{L}-\bell)}{C^{\text{total}}_{|\Vec{L}-\bell|}} f^\phi(\bell,\Vec{L}-\bell) \, .
\end{equation}
We then replace $C^{\text{total}}_{|\Vec{L}-\bell|} \rightarrow C^{\text{total}}_\ell$ and expand the \emph{asymmetric} lensing response function in Fourier series
\begin{equation}\label{mulresasym}
    f^\phi(\bell,\Vec{L}-\bell) = \sum_{m \geq 0} \tilde{f}^m_{L,\ell}\cos(m\theta_{\vec{L},\bell})\, , 
\end{equation}
which now involves both even and odd multipoles. In the squeezed limit, the expansion is still dominated by $m=0$ and $m=2$ and the expressions for $\tilde{f}^0_{L,\ell}$ and $\tilde{f}^2_{L,\ell}$ reduce to their symmetric counterparts $f^0_{L,\ell}$ and $f^2_{L,\ell}$, respectively. We expect the majority of the signal-to-noise to remain in these multipoles even for smaller-scale lenses (as was the case for the symmetric estimator above). The argument for foreground immunity of the $m=2$ estimator still holds in this asymmetric case, also (see Sec.~\ref{subsec:foregrounds}).

We end this section by noting that an alternative to considering the $m=2$ estimator (plus higher-multipole estimators) is to remove the $m=0$ contribution from the standard QE, as was done in~\cite{Fabbian2019}. However, it is difficult to find an efficient implementation of this scheme since it requires a very accurate separable approximation to the monopole estimator as any error in the monopole removal will cause leakage of foreground contamination.
 

\subsection{Foreground immunity}
\label{subsec:foregrounds}

To see why extragalactic foregrounds predominantly affect the $m=0$ multipole estimator, we consider a simple toy model of extragalactic sources all with the same circularly symmetric angular profile $F(\vec{x})$ in the temperature map. Assume that these are Poisson sampled from a fluctuating-mean source density $n(\vec{x}) = \bar{n}[1+b\delta(\vec{x})]$ where $\bar{n}$ is the global mean source density, $\delta(\vec{x})$ is a projected density field correlated with the CMB lensing potential, and $b$ is a linear bias. If we average over the Poisson fluctuations at fixed $\delta(\vec{x})$, the two-point function of the source contribution to the temperature map, $f(\vec{x})$, satisfies
\begin{equation}
\langle f(\bell_1) f(\bell_2) \rangle_{\text{Poisson}} = \bar{n} F(\bell_1) F(\bell_2) \left[ (2\pi)^2 \delta^{(2)}(\bell_1+\bell_2) + b \delta(\bell_1+\bell_2) \right] 
\end{equation}
to first order in $\delta$ and for $\bell_1 \neq 0$ and $\bell_2\neq 0$. If we consider applying a multipole estimator $\hat{\phi}^m(\vec{L})$ to a map composed of CMB, instrument noise and foregrounds $f$, the foreground bias in the correlation of the reconstructed $\phi$ with the true $\phi$ is $\langle \hat{\phi}^m_{\vec{L}}(f,f) \phi(\vec{L}')\rangle$, where $\hat{\phi}^m_{\vec{L}}(f,f)$ is the multipole estimator applied to the field $f$. At leading order, and for $\vec{L}\neq 0$, this bias is
\begin{align}
\langle \hat{\phi}^m_{\vec{L}}(f,f) \phi(\vec{L}')\rangle &=
\langle \delta(\vec{L}) \phi(\vec{L}') \rangle \frac{A_L^m}{2} \int \frac{d^2 \bell}{(2\pi)^2}\, g^m_{L,\ell} \cos (m \theta_{\Vec{L},\bell}) b \bar{n} 
F(\bell+\Vec{L}/2) F(\Vec{L}/2-\bell) \nonumber \\
&=
\langle \delta(\vec{L}) \phi(\vec{L}') \rangle \frac{A_L^m}{2} \int \frac{d^2 \bell}{(2\pi)^2}\, g^m_{L,\ell} \cos (m \theta_{\Vec{L},\bell}) b \bar{n} 
[F(\ell)]^2 \left[1+ \mathcal{O}(x^2)\right] ,
\label{eq:primbispecbias}
\end{align}
so the bias predominantly appears in the $m=0$ estimator. Note that this holds true even at relatively large $L$ provided that the source profile $F(\vec{x})$ is very compact. It also remains true if we replace $T(\bell+\Vec{L}/2) T(\Vec{L}/2-\bell)$ with $T(\bell) T(\Vec{L}-\bell)$ to allow fast evaluation of the estimator, as discussed above. In this case, Eq.~\eqref{eq:primbispecbias} becomes
\begin{align}
\langle \hat{\phi}^m_{\vec{L}}(f,f) \phi(\vec{L}')\rangle &\rightarrow
\langle \delta(\vec{L}) \phi(\vec{L}') \rangle \frac{A_L^m}{2} \int \frac{d^2 \bell}{(2\pi)^2}\, g^m_{L,\ell} \cos (m \theta_{\Vec{L},\bell}) b \bar{n} 
F(\bell) F(\Vec{L}-\bell) \nonumber \\
&=
\langle \delta(\vec{L}) \phi(\vec{L}') \rangle \frac{A_L^m}{2} \int \frac{d^2 \bell}{(2\pi)^2}\, g^m_{L,\ell} \cos (m \theta_{\Vec{L},\bell}) b \bar{n} 
[F(\ell)]^2 \left[1+ \mathcal{O}(x)\right] ,
\end{align}
Although the expansion of $F(\bell)F(\vec{L}-\bell)$ now introduces terms suppressed by only one power of $x=L/\ell$, these have $m=1$ angular dependence and so do not bias the $m=2$ estimator. The $m=2$ foreground terms are still suppressed in the integrand by $(L/\ell)^2$.

If instead, we consider the power spectrum of the reconstructed lensing potential, foreground biases similar to those above arise from terms like
$\langle \hat{\phi}^m_{\vec{L}}(f,f) \hat{\phi}^m_{\vec{L}'}(T,T) \rangle$ since averaging over the unlensed CMB reduces the second quadratic estimator to $\phi(\vec{L}')$. These terms are similarly suppressed for $m \neq 0$. However, additional ``secondary bispectrum'' terms, of the form $\langle \hat{\phi}^m_{\vec{L}}(T,f) \hat{\phi}^m_{\vec{L}'}(T,f) \rangle$, where the CMB and Poisson fluctuations are averaged across quadratic estimators, are not suppressed. However, these biases are generally small; see~\cite{Schaan2019} and Sec.~\ref{test}. Foreground trispectrum biases also arise that involve the connected four-point function of the foregrounds $\langle \hat{\phi}^m_{\vec{L}}(f,f) \hat{\phi}^m_{\vec{L}'}(f,f) \rangle_c$. These are also suppressed in the limit where shot noise dominates. In this case, our toy model gives
\begin{equation}
\langle f(\bell_1) f(\bell_2) f(\bell_3) f(\bell_4) \rangle = (2\pi)^2 \bar{n} F(\bell_1) F(\bell_2) F(\bell_3) F(\bell_4) \delta^{(2)}(\bell_1+\bell_2+\bell_3+\bell_4) \, ,
\end{equation}
and the trispectrum bias separates to involve the same integral as in Eq.~\eqref{eq:primbispecbias}:
\begin{equation}
\langle \hat{\phi}^m_{\vec{L}}(f,f) \hat{\phi}^m_{\vec{L}'}(f,f) \rangle_c = (2\pi)^2 \delta^{2}(\vec{L}+\vec{L}') \bar{n} \left[
\frac{A_L^m}{2} \int \frac{d^2 \bell}{(2\pi)^2}\, g^m_{L,\ell} \cos (m \theta_{\Vec{L},\bell}) 
F(\bell+\Vec{L}/2) F(\Vec{L}/2-\bell) \right]^2 .
\end{equation}

\subsection{Full-sky formalism}
\label{sec:fullsky}

In this section, we construct the full-sky versions of the shear and $m=2$ estimators. We start with the real-space form of the shear estimator in the flat-sky limit, Eq.~\eqref{eq1111}. It is easy to extend this to the curved sky, using spherical-harmonic functions as a basis, converting the partial derivatives into covariant derivatives on the sphere $\partial \rightarrow \nabla$, and the integration measure from $d^2\vec{x}\rightarrow d^2 \hat{\vec{n}}$. These changes give
\begin{equation}
    \hat{\phi}^{\text{shear}}_{LM}=\frac{A^{\text{shear}}_L}{2}\int{d}^2\hat{\vec{n}}\, \left(\nabla^{\langle a}\nabla^{b\rangle}Y^{*}_{LM}\right)T(\hat{\vec{n}})\nabla_{\langle a}\nabla_{b\rangle}T^{F}(\hat{\vec{n}}) \, .
\label{eq:curvedshear}
\end{equation} 
This expression can be evaluated efficiently by writing the symmetric, trace-free derivatives of the spherical-harmonic functions in terms of spin-$\pm 2$ spherical harmonics and then using (fast) spherical-harmonic transforms; see Appendix~\ref{appB}. Evaluating the resulting Gaunt integral in terms of Wigner-$3j$ symbols allows us to derive the following harmonic-space form of the full-sky shear estimator:
%
\begin{equation}\label{curved1}
    \hat{\phi}^{\text{shear}}_{LM}=A^{\text{shear}}_L\sum_{\ell_1m_1}\sum_{\ell_2m_2}(-1)^Mg^{\text{shear}}_{\ell_1,\ell_2}(L)\begin{pmatrix}
\ell_1 & \ell_2 & L\\
m_1 & m_2 & -M
\end{pmatrix}T_{\ell_1m_1}T_{\ell_2m_2} \, .
\end{equation}
Here, the full-sky shear weight function $g^{\text{shear}}_{\ell_1,\ell_2}(L)$ has the same structure as its flat-sky counterpart
in Eq.~\eqref{eq.flatw}:
\begin{multline}
    g^{\text{shear}}_{\ell_1,\ell_2}(L)=\frac{1}{2}\sqrt{\frac{(2L+1)(2\ell_1+1)(2\ell_2+1)}{16\pi}}\frac{C^{TT}_{\ell_1}}{(C^{\text{total}}_{\ell_1})^2}\frac{d\ln C^{TT}_{\ell_1}}{d\ln \ell_1}\begin{pmatrix}
\ell_1 & \ell_2 & L\\
0 & 0 & 0
\end{pmatrix} \\ \times \omega^2_L
\left[\frac{\left(\omega^2_L+\omega^2_{\ell_1}-\omega^2_{\ell_2}\right)\left(\omega^2_L+\omega^2_{\ell_1}-\omega^2_{\ell_2}-2\right)}{2\omega^2_{\ell_1}\omega^2_{L}}-1\right] \, ,
\label{eq:gshearcurved}
\end{multline}
where $\omega_\ell^2\equiv\ell(\ell+1)$. The $3j$ symbol here enforces mode coupling, the spherical analogue of $\bell_1+\bell_2=\Vec{L}$, and the expression in square brackets accounts for $\cos (2\theta_{\Vec{L},\bell_1})$ noting that, by the cosine rule,
\begin{equation}
\cos\theta_{\Vec{L},\bell_1} = \frac{L^2 + \ell_1^2 - \ell_2^2}{2 L \ell_1} \qquad
\Rightarrow \qquad \cos(2\theta_{\Vec{L},\bell_1}) = \frac{\left(L^2 + \ell_1^2 - \ell_2^2\right)^2}{2 L^2 \ell_1^2} - 1 \, .
\end{equation}
With the usual curvature correction, $\ell \rightarrow \omega_\ell$, and assuming that all multipoles are much larger than $\sqrt{2}$, we recover the correspondence to Eq.~\eqref{eq.flat}.
The explicit form of the spherical normalization, $A^{\text{shear}}_L$, is given in Appendix~\ref{appA}.

One can similarly derive the full-sky version of the $m=2$ estimator, obtained by replacing $f^{\text{shear}}_{L,\ell}$ by $f^2_{L,\ell}$ (or the asymmetric form $\tilde{f}^2_{L,\ell}$) in the reconstruction weight function. We now have
%
\begin{equation}
    \hat{\phi}^{m=2}_{LM}=\frac{A^{m=2}_L}{2} \int{d}^2\hat{\vec{n}}\, \left(\nabla^{\langle a}\nabla^{b\rangle}Y^{*}_{LM}\right)T(\hat{\vec{n}})\nabla_{\langle a}\nabla_{b\rangle}T^{F,m=2}_L(\hat{\vec{n}}) \, ,
    \label{phishear1}
\end{equation} 
where the non-separable filtered temperature field $T^{F,m=2}_L(\hat{\vec{n}})$ is given by
\begin{equation}
    T^{F,m=2}_L(\hat{\vec{n}})=\frac{1}{\omega_L^2}\sum_{\ell m} \frac{2}{\omega_\ell^2(C^{\text{total}}_\ell)^2} f^2_{L,\ell}
  T_{\ell{m}}Y_{\ell{m}}(\hat{\vec{n}}) \, .
  \label{eq:nonsepmeqtwoweights}
\end{equation}
The harmonic-space form is
\begin{equation}\label{curvedm2}
    \hat{\phi}^{m=2}_{LM}=A^{m=2}_L\sum_{\ell_1m_1}\sum_{\ell_2m_2}(-1)^Mg^{m=2}_{\ell_1,\ell_2}(L)\begin{pmatrix}
\ell_1 & \ell_2 & L\\
m_1 & m_2 & -M
\end{pmatrix}T_{\ell_1m_1}T_{\ell_2m_2} \, ,
\end{equation}
where the weight function $g^{m=2}_{\ell_1,\ell_2}(L)$ is no longer a separable function of $L$, $\ell_1$ and $\ell_2$, and is given by
\begin{multline}
    g^{m=2}_{\ell_1,\ell_2}(L)=\sqrt{\frac{(2L+1)(2\ell_1+1)(2\ell_2+1)}{16\pi}}\frac{1}{(C^{\text{total}}_{\ell_1})^2}\begin{pmatrix}
\ell_1 & \ell_2 & L\\
0 & 0 & 0
\end{pmatrix} \\ \times
\left[\frac{\left(\omega^2_L+\omega^2_{\ell_1}-\omega^2_{\ell_2}\right)\left(\omega^2_L+\omega^2_{\ell_1}-\omega^2_{\ell_2}-2\right)}{2\omega^2_{\ell_1}\omega^2_{L}}-1\right] f^2_{L,\ell_1} \, .
\end{multline}

This $m=2$ estimator performs better than the shear estimator on small scales but is computationally expensive to evaluate due to the non-separability of the weight function. Reconstruction at each $L$ requires evaluation of a separate filtered temperature field, $T^{F,m=2}_L(\hat{\vec{n}})$, making the entire reconstruction a factor of $L_{\text{max}}$ slower than for the separable shear estimator. The reconstruction would therefore scale as $\mathcal{O}(L^4_\text{max})$, which is generally infeasible, particularly as the reconstruction typically has to be run many times for simulation-based removal of biases in the reconstructed lensing power spectrum. In Sec.~\ref{svd}, we suggest a simple way of writing the non-separable $f^2_{L,\ell}$ (or, actually, $\tilde{f}_{L,\ell}^2$) as a sum of separable terms, allowing reconstruction to be performed with far fewer filtered fields.

Finally, for completeness, we point out that the same prescription can be extended to calculate higher-multipole estimators even though, as noted above, most of the signal-to-noise is contained in the monopole and quadrupole. As an example, consider the $m=4$ estimator. On the flat sky, the real-space form of the estimator follows from noting that
\begin{equation}
\cos (4\theta_{\Vec{L},\bell}) = 8 \hat{L}_{\langle i} \hat{L}_j \hat{L}_k \hat{L}_{l\rangle} \hat{\ell}^{\langle{i}} \hat{\ell}^j \hat{\ell}^k \hat{\ell}^{l\rangle}\, .
\end{equation}
This allows us to express the $m=4$ response function directly in terms of $\bell$ and $\Vec{L}$ resulting in the following real-space estimator on the full sky (for $L \geq 4$):
\begin{equation}\label{m4}
    \hat{\phi}^{m=4}_{LM}=\frac{A^{m=4}_L}{2} \int{d}^2\hat{\vec{n}}\, \left(\nabla^{\langle a} \nabla^b \nabla^c
    \nabla^{d\rangle}Y^{*}_{LM}\right)T(\hat{\vec{n}})\nabla_{\langle a}\nabla_b \nabla_c \nabla_{d\rangle}T^{F,m=4}_L(\hat{\vec{n}}) \, ,
\end{equation}
where the filtered field is now
\begin{equation}
    T^{F,m=4}_L(\hat{n})=\frac{8}{\omega_L^4} \sum_{\ell m} \frac{1}{\omega_\ell^4(C^{\text{total}}_\ell)^2} f^4_{L,\ell}
  T_{\ell m}Y_{\ell m}(\hat{\vec{n}}) \, .
\end{equation}
Here, we have simply replaced the flat-sky $\ell^4$, which comes from converting multiplication by the unit vector $\hat{\bell}$ to a derivative, with $\omega_\ell^4$ and similarly for $L^4$. The harmonic-space form is
\begin{equation}\label{curvedm4}
    \hat{\phi}^{m=4}_{LM}=A^{m=4}_L\sum_{\ell_1m_1}\sum_{\ell_2m_2}(-1)^Mg^{m=4}_{\ell_1,\ell_2}(L)\begin{pmatrix}
\ell_1 & \ell_2 & L\\
m_1 & m_2 & -M
\end{pmatrix}T_{\ell_1m_1}T_{\ell_2m_2} ,
\end{equation}
with weight function 
\begin{multline}
    g^{m=4}_{\ell_1,\ell_2}(L)=\sqrt{\frac{(L+4)!}{(L-4)!}}\sqrt{\frac{(\ell_1+4)!}{(\ell_1-4)!}}\sqrt{\frac{(2L+1)(2\ell_1+1)(2\ell_2+1)}{16\pi}}\begin{pmatrix}
\ell_1 & \ell_2 & L\\
4 & 0 & -4
\end{pmatrix}\frac{1}{2}\left[1+(-1)\right]^{\ell_1+\ell_2+L} \\
\times \frac{1}{\omega_L^4 \omega_{\ell_1}^4} \frac{1}{\left(C^{\text{total}}_{\ell_1}\right)^2} f^4_{L,\ell} \, .
\end{multline}
Note that in going from the flat-sky to full-sky, we could have chosen to use $\sqrt{(\ell+4)!/(\ell-4)!}$ instead of $\omega_\ell^4$ for the terms that arise with the 4th derivatives. The fractional differences are $\mathcal{O}(1/\ell^2)$ and would produce only small changes in optimality (at intermediate and small scales) while simplifying the above weight functions.

We test the full-sky shear estimator using full-sky lensed CMB simulations with the specifications of an upcoming Stage-3 experiment, with $1.4\,\text{arcmin}$ beam width (full-width at half maximum) and $7\,\mu\text{K-arcmin}$ instrument white noise at a frequency of $148\,\text{GHz}$. We first test the estimator using full-sky temperature maps without foreground contamination. The cross-power spectrum between the reconstructed convergence field $\kappa$ (related to the lensing field via $\kappa=-\nabla^2\phi/2$) and the input agrees with the theory lensing spectrum at the percent level. This shows that our estimator is correctly normalised (we use the full-sky normalisation given in Appendix~\ref{appA}). We further verify that we can recover an unbiased estimate of 
the convergence power spectrum, $C^{\kappa\kappa}_L = L^2 (L+1)^2 C_L^{\phi\phi}/4$, by subtracting several biases from the empirical auto-spectrum of the reconstruction, $C^{\hat{\kappa}\hat{\kappa}}_L$:
\begin{equation}
    \hat{C}^{\kappa\kappa}_L=C^{\hat{\kappa}\hat{\kappa}}_L-N^{(0)}_L-N^{(1)}_L .
\end{equation}

\begin{figure}[hbt!]
\includegraphics[width=0.7\columnwidth]{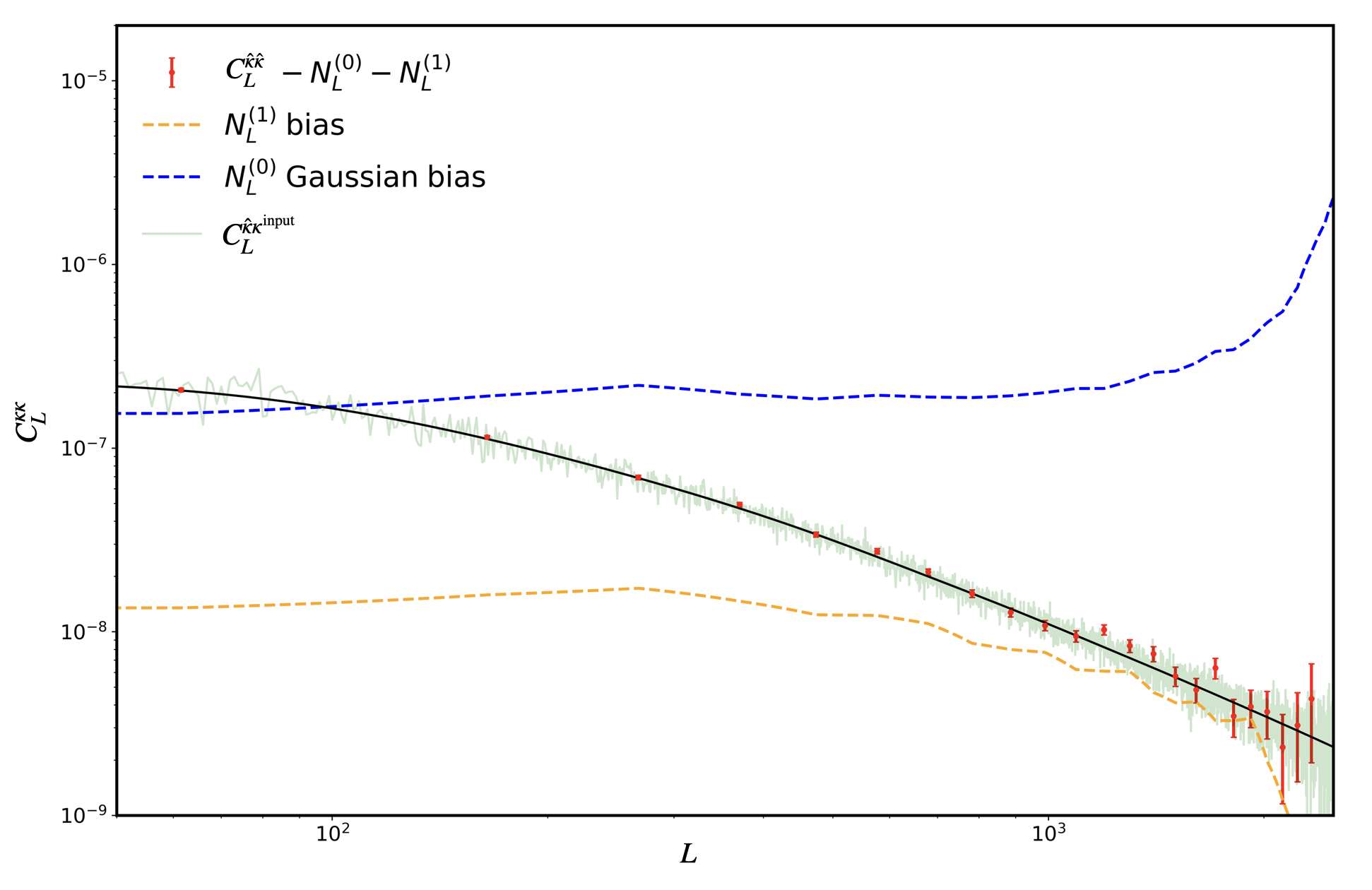}
\centering
\caption{\label{fig:reconstruct} Lensing convergence reconstructed on the full sky with the shear-only estimator (Eq.~\ref{eq:curvedshear}) applied to a simulated, foreground-free, noisy CMB temperature map. The cross-power spectrum of the reconstructed and input convergence is shown in green, and bandpowers with $\Delta{L}=60$  of the auto-power spectrum of the reconstruction are in red. The latter is corrected for $N^{(0)}$ and $N^{(1)}$ bias terms (shown in blue and orange, dashed, respectively). Both the auto- and cross-spectra agree well with the expected power (black).
}
\end{figure}

Here, $N^{(0)}_L$ is the Gaussian bias produced from the disconnected part of the CMB four-point function that enters $\langle C^{\hat{\kappa}\hat{\kappa}}_L\rangle$. This can be thought of as the power spectrum of the statistical reconstruction noise sourced by chance, Gaussian fluctuations in the CMB and instrument noise that mimic the effects of lensing.
We estimate this bias by forming different pairings of the simulation that is being treated as data and independent simulations following the method described in \cite{Namikawa2013}. We use 100 different realisations of simulation pairs to obtain an average over simulations. The $N^{(1)}$ bias, which arises from the connected part of the CMB four-point function and at leading order is linear in the lensing power spectrum~\cite{Kesden_2003}, is estimated using 200 pairs of simulations, with the same lensing realisation for each member of the pair but different unlensed CMB realisations, based on \cite{Story2015}.  The de-biased bandpowers of the shear reconstruction are shown in Fig.~\ref{fig:reconstruct}.

\section{Beyond the large-scale-lens regime: SVD expansion of the multipole kernels}\label{svd}

\begin{figure}[t]
\centering
\includegraphics[width=8.8cm]{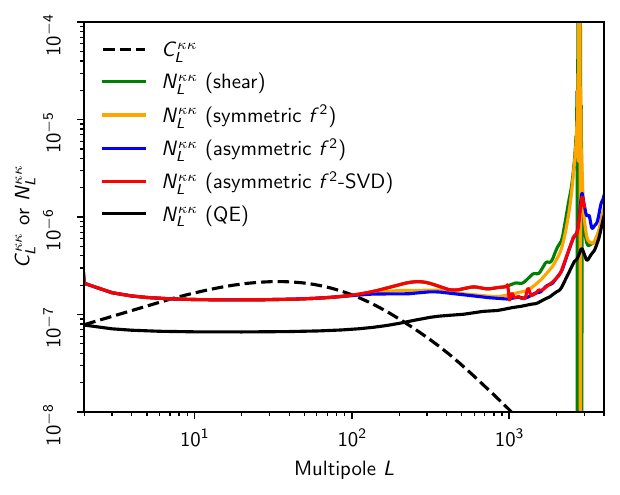}
\caption{\label{fig:noise} Lensing (convergence) reconstruction noise power spectra for the standard quadratic estimator (QE; black), the shear estimator (green), the $m=2$ symmetric estimator (symmetric $f^2$; orange), the $m=2$ asymmetric estimator (asymmetric $f^2$; blue), and the SVD approximation to the latter (asymmetric $f^2$-SVD; red). The lensing convergence power spectrum is also plotted (dashed black). Note that the convergence spectra are related to the spectra of the lensing potential by, e.g., $N_L^{\kappa\kappa} = L^2(L+1)^2 N_L^{\phi\phi}/4$. The survey specifications are the same as in Sec.~\ref{sec:fullsky}.}
\end{figure}

The shear estimator in Eq.~\eqref{eq:curvedshear} is separable and so efficient to evaluate, but this comes at the cost of increased noise in the reconstruction on small scales. The sub-optimality of the shear estimator is apparent from Fig.~\ref{fig:noise}, which shows that its disconnected noise bias $N_L^{(0)}$ has a spike at small scales (see also Fig.~2 in~\cite{Schaan2019}). This spike arises because the shear estimator has zero response to lenses at this particular scale. The noise biases in the figure are computed on the full sky using Eq.~\eqref{eq:fullnoise}.

The full $m=2$ estimator, which in this paper we approximate by Eq.~\eqref{phishear1} with non-separable weights~\eqref{eq:nonsepmeqtwoweights}, has better noise performance than the shear estimator for $L>100$ (for the survey specifications adopted here) as shown in Fig.~\ref{fig:noise}. In particular, if the weights are constructed from the $m=2$ component of the \emph{asymmetric} lensing response function, $\tilde{f}^2_{L,\ell}$, the noise spike is eliminated. However, such $m=2$ estimators are inefficient to evaluate since the weights are not separable.



A simple work-around, which we have found to perform quite well, is to retain the squeezed-limit, separable approximation (i.e., the shear estimator) on large scales where its performance is similar to the full $m=2$ estimator, but to approximate the $\tilde{f}^2_{L,\ell}$ as a sum of separable terms on smaller scales. We find these separable terms by singular-value decomposition (SVD) \cite{SVD}. 


In detail, we construct a hybrid approximation to $\tilde{f}^2_{L,\ell}$ as follows. For $L < L_\ast$, we use the shear approximation; for $L > L_\ast$, we perform a singular-value decomposition of the block of the asymmetric, \emph{convergence} response function, $2 \tilde{f}^2_{L,\ell} / \omega_L^2$, with $L_\ast \leq L \leq L_{\text{max}}$ and $2 \leq \ell \leq \ell_{\text{max}}$, and approximate this with the first $n$ largest singular values. We found that keeping $n=20$ SVD terms gives a reasonable balance between computational efficiency and optimality. We chose the lensing multipole $L_\ast$ at which to switch such that the reconstruction noise on the SVD-based estimator is lower than that of the shear, which for our survey parameters is $L_\ast = 1000$. (Note that there is a range of $L$ in which the above condition is true, and $L_\ast$ was chosen empirically based on good SVD convergence. Other metrics could certainly be used to determine $n$ and $L_\ast$.) The complete form of our hybrid-SVD response function is
\begin{equation}
    \frac{2\tilde{f}^{\text{hybrid-SVD}}_{{L},\ell}}{\omega_L^2} = \begin{cases} C^{TT}_\ell d\ln C^{TT}_\ell / d\ln{\ell} \, , & L<L_\ast  \\ \sum^n_{i=1}\Lambda_i U_{L,i} V_{\ell,i} \, & L\geq L_\ast \, . \end{cases}
\end{equation}
Here $\Lambda_i$ corresponds to the $i$th  singular value, $U_{L,i}$ are the components of the $i$th left singular vector and $V_{\ell,i}$ are the components of the $i$th right singular vector. We see that the SVD naturally decomposes the response into a sum of separable terms and hence the reconstruction can be performed efficiently for each component. The separable SVD estimator is given explicitly by (for $L \geq L_\ast$)
\begin{equation}
    \hat{\phi}^{\text{SVD}}_{LM}=\frac{A_L^{\text{SVD}}}{2}\sum_{i=1}^n \Lambda_i U_{L,i}
    \int{d}^2\hat{\vec{n}}\, \left(\nabla^{\langle a}\nabla^{b\rangle}Y^{*}_{LM}\right)T(\hat{\vec{n}})\nabla_{\langle a}\nabla_{b\rangle}T^{F,\text{SVD}}_i(\hat{\vec{n}}) \, ,
\end{equation}
where the filtered field for the $i$th singular value is
\begin{equation}
 T^{F,\text{SVD}}_i(\hat{\vec{n}})=\sum_{\ell m} \frac{V_{\ell,i}}{\omega_\ell^2(C^{\text{total}}_\ell)^2}
  T_{\ell{m}}Y_{\ell{m}}(\hat{\vec{n}}) \, .
\end{equation}

The normalisation $A_L^{\text{SVD}}$ is chosen, as usual, to ensure the estimator has the correct response to lensing. We show in Sec.~\ref{test} that the separable approximation is still very effective at suppressing foregrounds, as expected since we have not altered the $m=2$ geometric structure of the estimator. As can be seen from Fig.~\ref{fig:noise}, the noise performance is very close to that of the full (asymmetric) $m=2$ estimator for $L \geq  L_\ast$, and around a factor of two better than the shear estimator.

\section{Testing the sensitivity to foregrounds using simulations}\label{test}

\begin{figure}[ht]
\centering
\includegraphics[width=0.5\columnwidth]{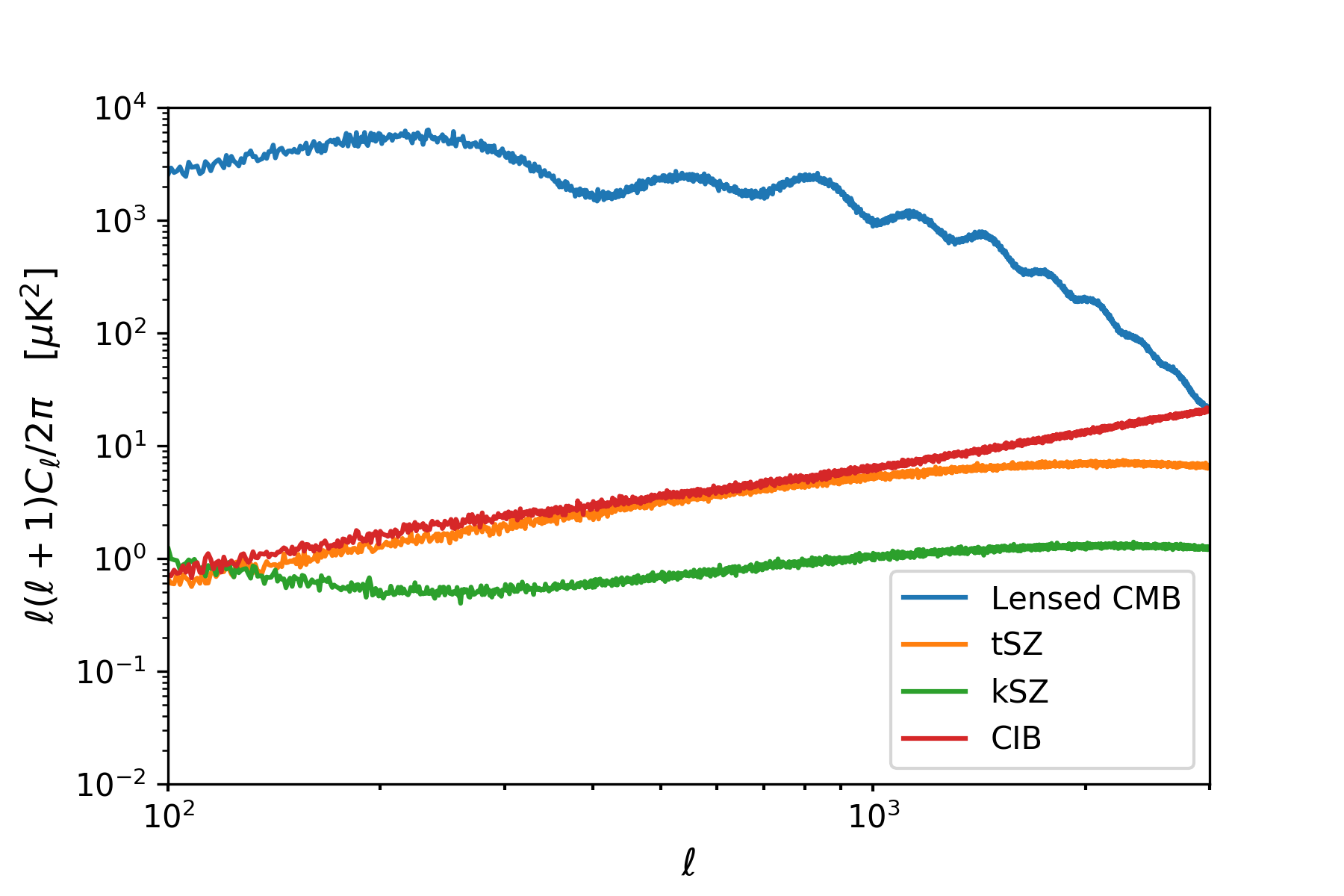}
\caption{\label{fig:epsart} Power spectra of the tSZ (orange), kSZ (green) and CIB (red) extragalactic foregrounds at $148\,\text{GHz}$ from the Websky simulation~\cite{Stein2020}. The lensed CMB power spectrum is also shown (in blue).
The combined total map is matched-filtered and masked for point sources above a flux density of $5\,\text{mJy}$ in our tests. }
\label{foreground}
\end{figure}

To test the sensitivity of the estimators to extragalactic foregrounds, we use the component maps of the Websky extragalactic foreground simulations~\cite{Stein2020}. These include CIB, tSZ and kSZ at $143\,\text{GHz}$. The power spectra of these foregrounds are shown in Fig.~\ref{foreground}. In a real analysis, bright galaxy clusters and sources would be dealt with either by masking (i.e., excising regions around them) or in-painting (masking, but with the resulting holes filled with constrained realizations). We mimic this for point sources in our analysis, without introducing the complications of having to deal with masked or in-painted maps, as follows. We apply a matched-filter, with the profile corresponding to the instrumental beam and noise power given by the sum of instrumental noise and foreground power, to maps including the full lensed CMB plus foregrounds plus white noise. Sources with recovered flux density greater than $5\,\text{mJy}$ are catalogued and regions around them are removed from the foreground maps only. These masked foreground maps are then combined with lensed CMB and $7\,\mu\text{K-arcmin}$ white noise to form the final temperature map given by $T_{\text{total}}=T_{\text{CMB}}+T_{f}+T_{\text{noise}}$, where we have written explicitly the contributions to the observed temperature map from the lensed CMB $T_{\text{CMB}}$, the extragalactic foregrounds $T_f$ and the detector noise $T_{\text{noise}}$. We do not mimic the masking of bright galaxy clusters. As noted below, this means that our results for the bias should be considered rather extreme, particularly for the trispectrum bias.

To assess the bias induced by foregrounds in the auto-power spectrum of the reconstruction, $\hat{C}^{\kappa\kappa}_L$, we evaluate the primary foreground bispectrum $2\langle\mathcal{Q}[T_f,T_f]\kappa\rangle$ and the foreground trispectrum term $\langle\mathcal{Q}[T_f,T_f]\mathcal{Q}[T_f,T_f]\rangle_c$, from which the disconnected (Gaussian) contribution is subtracted using simulations. Here $\mathcal{Q}[T_A,T_B]$ represents a quadratic estimator (we consider the standard quadratic, shear and hybrid-SVD estimators) applied to maps $T_A$ and $T_B$. We do not consider the secondary bispectrum bias discussed in \cite{Schaan2019}, as it was found to be subdominant to the primary bispectrum and the trispectrum biases (and we expect the same to hold for our estimator variants).

\begin{figure}[H]
\centering
\includegraphics[width=1\columnwidth]{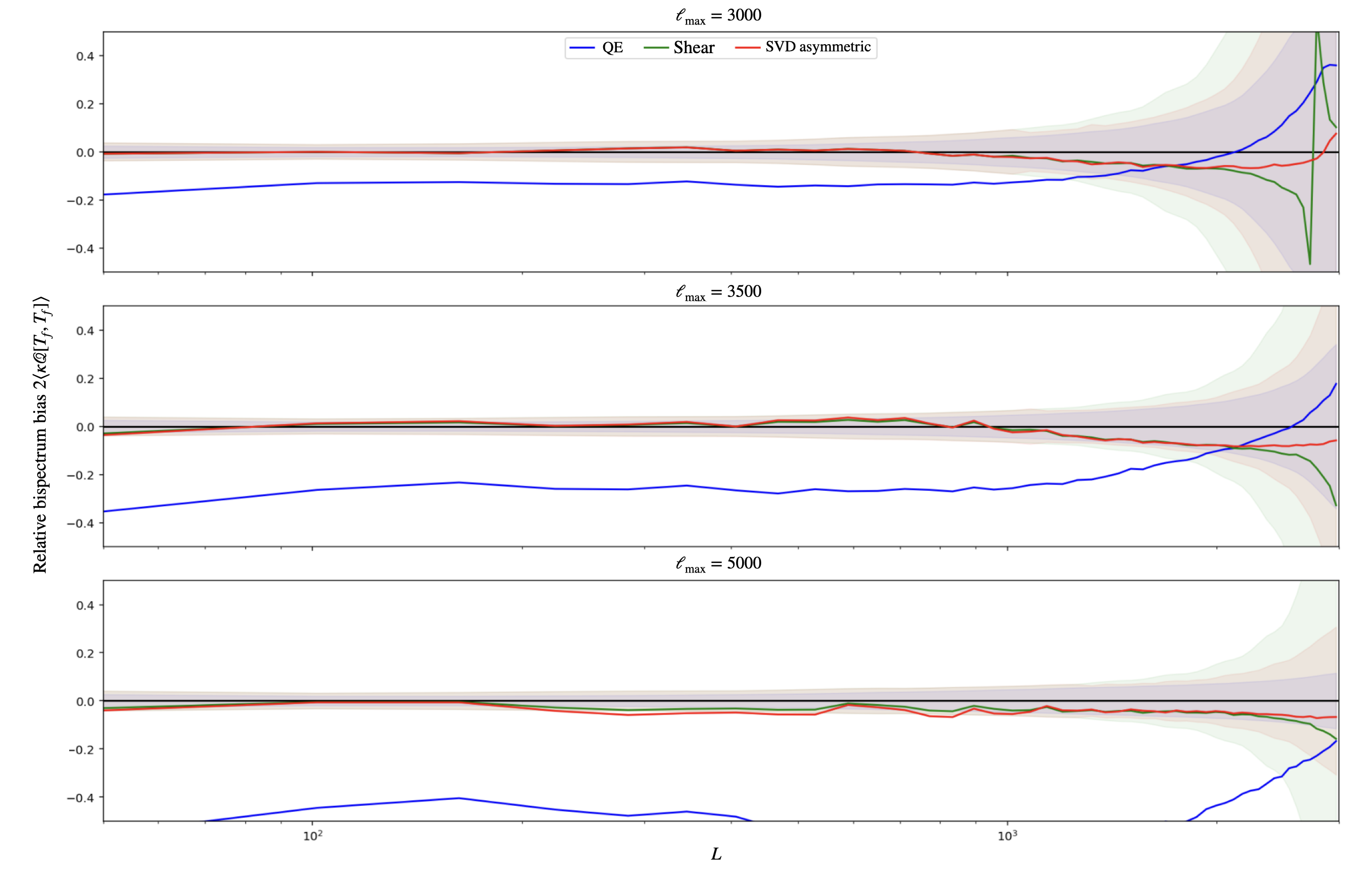}
\caption{\label{fig:bi} Relative primary bispectrum bias on the CMB convergence power spectrum due to the combined effect of foregrounds for the standard quadratic estimator (QE; blue), the shear estimator (green) and the hybrid-SVD estimator based on SVD of the asymmetric $\tilde{f}^2_{L,\ell}$ response (SVD asymmetric; red). From top to bottom, we vary the maximum CMB temperature multipole used in the reconstruction, $\ell_{\text{max}}=\{3000,3500,5000\}$. The shaded bands indicate the $1\,\sigma$ statistical reconstruction error for the different estimators for bandpowers of width $\Delta L = 60$. 
The bias is well contained within the statistical errors for the shear and hybrid-SVD estimators but there is significant bias in the standard quadratic estimator.}
\end{figure}

\begin{figure}[H]
\centering
\includegraphics[width=1\columnwidth]{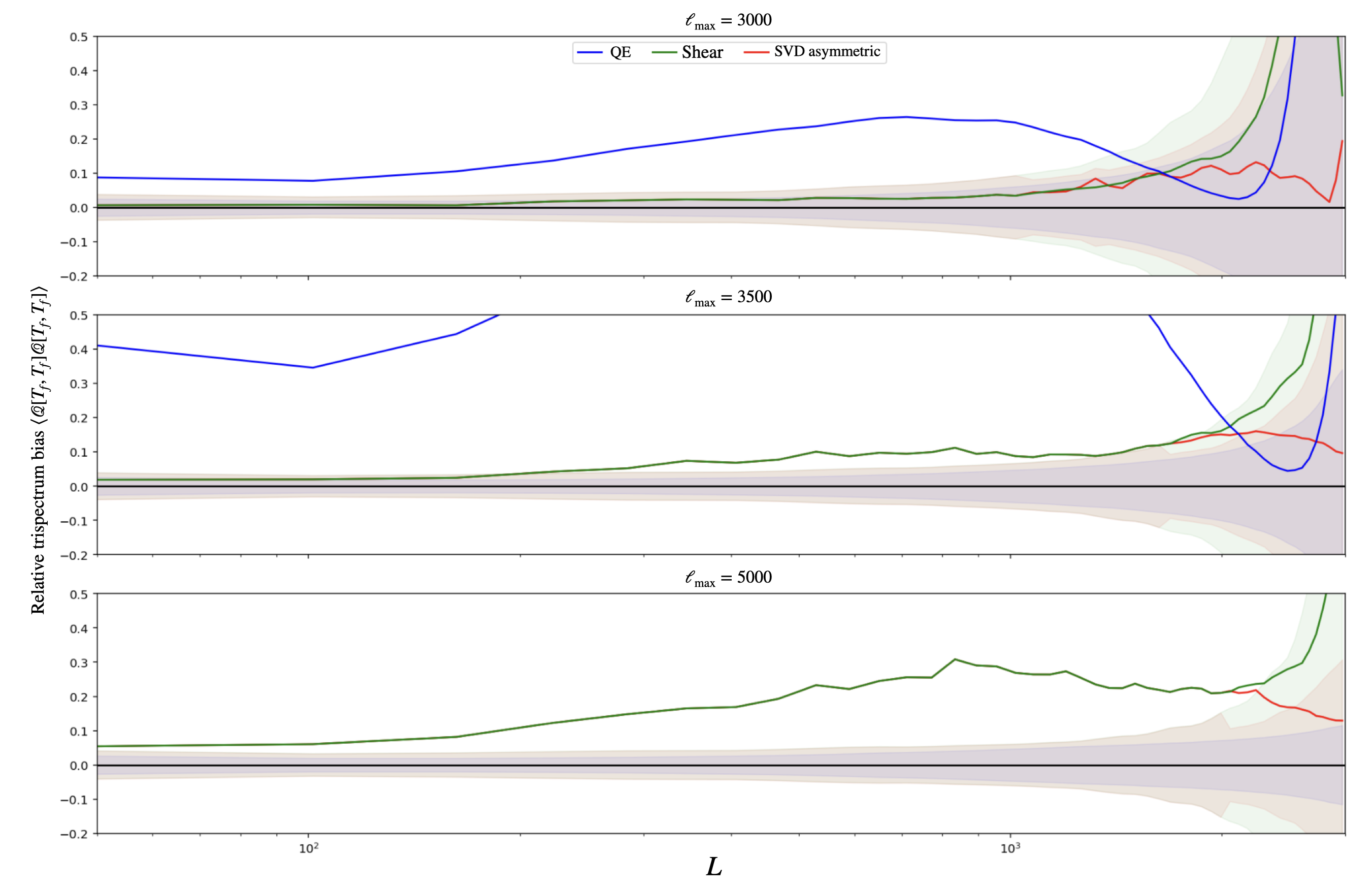}
\caption{\label{fig:tri} As Fig.~\ref{fig:bi} but for the relative trispectrum bias. Similar to the bispectrum biases, the shear and hybrid-SVD estimators suppress the trispectrum bias significantly across most of the lensing multipole range. Note that the bias for the standard quadratic estimator with $\ell_{\text{max}}=5000$ is off the scale of the plot. It is worth noting that the trispectrum bias is particularly sensitive to bright tSZ clusters and further improvement can be obtained via cluster masking. 
}
\end{figure}

The primary bispectrum bias on the lensing power spectrum can be seen in Fig.~\ref{fig:bi} for three choices of the maximum CMB multipole used in the reconstruction: $\ell_{\text{max}}=3000$, $3500$ and $5000$. Power spectra are binned with $\Delta{L}=60$. For all the $\ell_{\text{max}}$ choices, significant biases are observed in the standard quadratic estimator, while the shear estimator can remove the bias very effectively, in agreement with the flat-sky results of \cite{Schaan2019}. Furthermore, one can see that the bias induced in the hybrid-SVD estimator is smaller than that of the shear on small scales and comparable to that of the shear on large scales. The improvement in noise performance of the hybrid-SVD estimator compared to the shear estimator can also be appreciated, where the shaded $1\,\sigma$ bandpower errors (which include reconstruction noise and lensing sample variance) for the hybrid-SVD estimator (red) lie between the shear (green) and the standard quadratic estimator (blue), the latter having the lowest variance but a large foreground bias.
Similar improvements can be seen in Fig.~\ref{fig:tri}, where the trispectrum bias reduces significantly when switching from the standard quadratic estimator to the shear or hybrid-SVD estimator. Although for $\ell_{\text{max}}=3500$ and $\ell_{\text{max}}=5000$, the bias is no longer significantly smaller than the statistical error, the improvement compared with the standard quadratic estimator is still large. Furthermore, it should be noted that the trispectrum bias is particularly sensitive to the most massive galaxy clusters, which can be straightforwardly detected and removed by masking or inpainting in a real analysis. As we have not carried this out here, we expect a significant reduction in trispectrum bias for all estimators in practice.

\section{Discussion and Conclusions}

We showed how to formulate foreground-immune multipole estimators for CMB lensing reconstruction, particularly the $m=2$ estimator that contains most of the signal-to-noise, on the spherical sky. This allows the straightforward application of the estimators proposed in~\cite{Schaan2019} to large-area surveys such as Planck, AdvACT and the forthcoming Simons Observatory. Generally, these estimators are not separable and so cannot easily be evaluated efficiently. Previous separable approximations -- the shear estimator introduced in~\cite{Schaan2019} -- have sub-optimal reconstruction noise when reconstructing small-scale lenses. We presented a simple, first attempt at producing a separable approximation to the full $m=2$ estimator based on singular-value decomposition of the part of its response function at intermediate and large lensing multipoles. We tested the performance of this hybrid-SVD estimator, along with the shear approximation and the standard quadratic estimator, on the Websky~\cite{Stein2020} foreground simulation. As in the flat-sky tests considered in~\cite{Schaan2019}, we found the shear estimator to be very effective in suppressing foreground biases even on single-frequency maps. The same is true of the hybrid-SVD estimator, but it has the advantage of higher signal-to-noise on small scales.

The field of CMB lensing has experienced a fast transition from first detection~\cite{Smith:2007rg,Das_2011} to precision measurements in the last 15 years. With current and upcoming surveys of the CMB, such as AdvACT, SPT-3G and Simons Observatory, probing the millimetre sky with increasing resolution and sky coverage, we can expect further rapid improvements in the quality of lensing products reconstructed from the CMB. However, improvements in statistical noise must be met with more stringent control of systematic effects, such as those from extragalactic foregrounds in temperature maps. The methods explored in this paper provide a robust way to measure CMB lensing, which is largely immune to the effect of these foregrounds. They can be added to the existing repertoire of methods to mitigate foregrounds, such as multi-frequency cleaning and bias hardening, and can be used in combination with these to improve optimality further. For example, it was found in~\cite{darwish2021optimizing} that a robust estimator to reduce foreground biases while having a low impact on signal-to-noise tends to consist of a combination of bias hardening (for point sources and tSZ cluster profiles), explicit tSZ deprojection in multi-frequency foreground cleaning \emph{and} the shear estimator. \fr{Furthermore, Ref.~\cite{PhysRevD.106.063534} pointed out that extragalactic foregrounds in temperature can also bias $B$-mode delensing efforts for upcoming experiments like Simons Observatory, leading to bias in the inferred amplitude of the power spectrum of primordial gravitational waves if the non-Gaussianity of extragalactic foregrounds is not accounted for. These biases can also be effectively mitigated without significantly compromising the delensing efficiency if the lensing map is obtained using the shear-only estimator described here.}

\fr{In this work, we did not explore the impact of polarized extragalactic foregrounds. This is because for current and next-generation surveys, like ACT and Simons Observatory, the temperature-based ($TT$) estimator carries significant weight (about $2/3$ of the statistical weight for an ACT-like survey) and the foreground bias in temperature-based reconstruction is larger than for polarization since the degree of polarization of extragalactic foregounds is small compared to the CMB fluctuations. However, for future deep surveys, such as CMB-S4, the polarization channel will play a dominant role in lensing reconstruction when noise and foreground levels become below the lensing $B$-mode levels of around $5\,\mu\text{K-arcmin}$~\cite{PhysRevD.105.123519}, giving significant improvements in the signal-to-noise of the reconstruction. Although extragalactic-foreground biases to polarization-based lensing reconstructions are small compared to temperature, particularly for the $EB$ estimator, which is naturally a shear estimator and will carry most of the statistical weight for future surveys, the lower statistical reconstruction noise means the absolute requirements on foreground bias will be much more stringent. Extending the shear estimator for polarization-based estimators would enable a tighter control of potential polarized extragalactic-foreground contamination (coming mainly from bright radio and infrared point sources) as shown in \cite{PhysRevD.107.023504} for the flat-sky case. There it was found there that using the minimum-variance combination of shear estimators incurs only a modest $20\%$ noise penalty compared to the standard minimum-variance estimator for a CMB-S4-like survey. An extension of such polarization-based shear estimators to the full-sky case would be interesting but is
deferred to future work.}

\begin{acknowledgments}
We thank  William Coulton for providing the code for the matched-filter and Emmanuel Schaan and Simone Ferraro for useful discussions. This work used resources of the Niagara supercomputer at the SciNet HPC Consortium and the National Energy Research Scientific Computing Center. FQ acknowledges the support from a Cambridge Trust scholarship.
AC acknowledges support from the STFC (grant numbers ST/N000927/1 and ST/S000623/1). BDS acknowledges support from the European Research Council (ERC) under the European Union’s Horizon 2020 research and innovation programme (Grant agreement No. 851274) and from an STFC Ernest Rutherford Fellowship. 

\end{acknowledgments}

\appendix

\section{Full-sky normalization and $N^{(0)}_L$}\label{appA}

In this appendix we review the normalisation of full-sky quadratic estimators and the disconnected noise bias of their reconstructed power spectrum, following, e.g.,~\cite{Okamoto2003}.

We start with a general, full-sky quadratic estimator for $\phi$:
\begin{equation}
    \hat{\phi}_{LM}=A_L\sum_{\ell_1m_1}\sum_{\ell_2m_2}(-1)^M g_{\ell_1,\ell_2}(L)\begin{pmatrix}
\ell_1 & \ell_2 & L\\
m_1 & m_2 & -M
\end{pmatrix}T_{\ell_1m_1}T_{\ell_2m_2} \, .
\label{eq:appa1}
\end{equation}
The full-sky lensing response is
\begin{equation}
    \langle{T_{\ell_1 m_1}T_{\ell_2 m_2}}\rangle_{\text{CMB}}=(-1)^{m_1} C^{TT}_{\ell_1} \delta_{\ell_1\ell_2}\delta_{m_1-m_2}+\sum_{L M} (-1)^{M}\begin{pmatrix}
\ell_1 & \ell_2 & L\\
m_1 & m_2 & -M
\end{pmatrix}{f^\phi_{\ell_1 L \ell_2}}\phi_{LM} \, ,
\label{eq:appa2}
\end{equation}
where the weight $f^{\phi}_{\ell_1 L \ell_2}=C^{TT}_{\ell_1}{F_{\ell_2 L \ell_1}}+C^{TT}_{\ell_2}{F_{\ell_1 L \ell_2}}$ with
\begin{equation}
    F_{\ell_1 L \ell_2}=\left[L(L+1)+\ell_2(\ell_2+1)-\ell_1(\ell_1+1)\right]\sqrt{\frac{(2L+1)(2\ell_1+1)(2\ell_2+1)}{16\pi}}\begin{pmatrix}
\ell_1 & L& \ell_2\\
0 & 0 & 0
\end{pmatrix} \, .
\end{equation}
Note that $f^\phi_{\ell_1 L \ell_2}$ is symmetric in $\ell_1$ and $\ell_2$. In practice, we use the lensed power spectrum in $f^\phi_{\ell_1 L \ell_2}$, which is a good approximation to the true non-perturbative response~\cite{Lewis:2011fk}.

The normalisation $A_L$ is determined by demanding that the estimator is unbiased, i.e., $\langle \hat{\phi}_{LM} \rangle_{\text{CMB}} = \phi_{LM}$. Evaluating the average of Eq.~\eqref{eq:appa1} over the unlensed CMB fluctuations, the first term on the right of Eq.~\eqref{eq:appa2} only contributes at $L=0$ and so can be dropped. Simplifying the contribution of the second term with the properties of the $3j$-symbols gives the normalisation as
\begin{equation}
\frac{1}{A_L} = \frac{1}{2L+1}\sum_{\ell_1 \ell_2} g_{\ell_1,\ell_2}(L) f^\phi_{\ell_1 L \ell_2} \, .    
\end{equation}

We now consider the disconnected (Gaussian) noise bias $N^{(0)}_L$ on the reconstructed power spectrum. We have
\begin{multline}
\langle \hat{\phi}_{LM} \hat{\phi}_{L'M'} \rangle_G = A_L A_{L'} \sum_{\ell_1m_1}\sum_{\ell_2m_2} \sum_{\ell'_1 m'_1}\sum_{\ell'_2 m'_2} (-1)^{M} (-1)^{M'} \begin{pmatrix}
\ell_1 & \ell_2 & L\\
m_1 & m_2 & -M \end{pmatrix} \begin{pmatrix}
\ell'_1 & \ell'_2 & L'\\
m'_1 & m'_2 & -M' \end{pmatrix} \\ \times g_{\ell_1,\ell_2}(L) g_{\ell'_1,\ell'_2}(L') \langle T_{\ell_1 m_1} T_{\ell_2 m_2} T_{\ell'_1 m'_1} T_{\ell'_2 m'_2} \rangle_G 
\, , \label{eq:appa3}
\end{multline}
where the subscript $G$ denotes the disconnected (Gaussian) part of the expectation value. For the CMB four-point function, we have
\begin{equation}
\langle T_{\ell_1 m_1} T_{\ell_2 m_2} T_{\ell'_1 m'_1} T_{\ell'_2 m'_2} \rangle_G = (-1)^{m_1} (-1)^{m_2} C_{\ell_1}^{\text{total}} C_{\ell_2}^{\text{total}} \left[ \delta_{\ell_1 \ell'_1} \delta_{m_1 -m'_1} \delta_{\ell_2 \ell'_2} \delta_{m_2 -m'_2} + \left(\ell'_1 m'_1 \leftrightarrow \ell'_2 m'_2 \right) \right] \, ,
\end{equation}
where we have dropped the contractions that couple the temperature fields within the same quadratic estimator as these only give $L=L'=0$ contributions. Substituting in Eq.~\eqref{eq:appa3}, and noting that parity enforces $\ell_1 + \ell_2 + L = \text{even}$ and $\ell'_1 + \ell'_2 + L' = \text{even}$, we have
\begin{equation}
 \langle \hat{\phi}_{LM} \hat{\phi}_{L'M'} \rangle_G  = (-1)^M \delta_{LL'} \delta_{M -M'} N_L^{(0)} \, ,   
\end{equation}
where
\begin{equation}
N_L^{(0)} = A_L^2 \sum_{\ell_1 \ell_2} C_{\ell_1}^{\text{total}} C_{\ell_2}^{\text{total}} g_{\ell_1,\ell_2}(L)
\left[g_{\ell_1,\ell_2}(L) + g_{\ell_2,\ell_1}(L) \right] \, .
\label{eq:fullnoise}
\end{equation}

\section{Full-sky shear estimator in harmonic space}\label{appB}

In this appendix we show how to write the full-sky shear estimator in Eq.~\eqref{eq:curvedshear},
\begin{equation}\label{BBB}
    \hat{\phi}^{\text{shear}}_{LM}=\frac{A^{\text{shear}}_L}{2}\int{d}^2\hat{\vec{n}}\, \left(\nabla^{\langle a}\nabla^{b\rangle}Y^{*}_{LM}\right)T(\hat{\vec{n}})\nabla_{\langle a}\nabla_{b\rangle}T^{F}(\hat{\vec{n}}) \, ,
\end{equation} 
in harmonic space as (Eq.~\ref{curved1})
\begin{equation}
\hat{\phi}^{\text{shear}}_{LM}=A^{\text{shear}}_L\sum_{\ell_1m_1}\sum_{\ell_2m_2}(-1)^Mg^{\text{shear}}_{\ell_1,\ell_2}(L)\begin{pmatrix}
\ell_1 & \ell_2 & L\\
m_1 & m_2 & -M
\end{pmatrix}T_{\ell_1m_1}T_{\ell_2m_2} \, ,
\label{eq:appB2}
\end{equation}
and determine the weight function $g^{\text{shear}}_{\ell_1,\ell_2}(L)$.

We start by converting the covariant derivatives on the sphere into expressions involving spin-weighted spherical harmonics (see~\cite{Goldberg:1966uu} and, e.g.,~\cite{Lewis:2001hp}). For $s>0$ derivatives, we have
\begin{align}
\nabla^{\langle a_1}\ldots \nabla^{a_s \rangle}Y_{\ell m}&=\left(-\frac{1}{2}\right)^s \left(m_+^{a_1} \ldots m_+^{a_s} \bar{\eth}^s Y_{\ell m} + m_-^{a_1} \ldots m_-^{a_s} \eth^s Y_{\ell m} \right) \nonumber \\
&= \left(-\frac{1}{2}\right)^s \sqrt{\frac{(\ell+s)!}{(\ell-s)!}}\left[(-1)^s m_+^{a_1} \ldots m_+^{a_s} {}_{-s}Y_{\ell m} + m_-^{a_1} \ldots m_-^{a_s} {}_s Y_{\ell m} \right] \, ,
\end{align}
where $\vec{m}_\pm \equiv \hat{\boldsymbol{\theta}} \pm i \hat{\boldsymbol{\phi}}$ are null basis vectors constructed from unit vectors along the coordinate directions of spherical-polar coordinates $(\theta,\phi)$.
Expanding the filtered field in \eqref{BBB} in terms of spherical harmonics, 
\begin{equation}
    T^{F}(\hat{\vec{n}})=\sum_{\ell m} \frac{1}{\omega_\ell^2} \frac{C^{TT}_\ell}{(C^{\text{total}}_\ell)^2}\frac{d\ln{C^{TT}_\ell}}{d\ln{\ell}}
    T_{\ell{m}}Y_{\ell{m}}(\hat{\vec{n}}) \, ,
\end{equation}
where we have used the flat-sky expression~\eqref{eq:flatshearfilter} with $\ell^2$ replaced by its usual spherical equivalent $\omega_\ell^2 = \ell(\ell+1)$, we have the contraction
\begin{equation}
\nabla^{\langle a_1}\ldots \nabla^{a_s \rangle}Y_{LM}^\ast \nabla_{\langle a_1}\ldots \nabla_{a_s \rangle}Y_{\ell_1 m_1} = \left(\frac{1}{2}\right)^s \sqrt{\frac{(L+s)!}{(L-s)!}}\sqrt{\frac{(\ell_1+s)!}{(\ell_1-s)!}} \left({}_{-s}Y_{LM}^\ast {}_{-s} Y_{\ell_1 m_1} + {}_s Y_{LM}^\ast {}_s Y_{\ell_1 m_1} \right) \, .
\end{equation}
Multiplying by $Y_{\ell_2 m_2}$ from the expansion of the unfiltered field in Eq.~\eqref{BBB}, the resulting integral over $\hat{\vec{n}}$ can be performed with the Gaunt integral to obtain
\begin{multline}
\int d^2 \hat{\vec{n}} \,   \left(\nabla^{\langle a_1}\ldots \nabla^{a_s \rangle}Y_{LM}^\ast \right) \left(\nabla_{\langle a_1}\ldots \nabla_{a_s \rangle}Y_{\ell_1 m_1}\right) Y_{\ell_2 m_2} = (-1)^M \left(-\frac{1}{2}\right)^s  \sqrt{\frac{(L+s)!}{(L-s)!}}\sqrt{\frac{(\ell_1+s)!}{(\ell_1-s)!}} \\ \times \sqrt{\frac{(2L+1)(2\ell_1+1)(2\ell_2+1)}{4\pi}} 
\begin{pmatrix}
\ell_1 & \ell_2 & L\\
m_1 & m_2 & -M
\end{pmatrix}
\begin{pmatrix}
\ell_1 & \ell_2 & L\\
s & 0 & -s
\end{pmatrix} \left[1+(-1)^{\ell_1+\ell_2+L}\right] \, ,
\end{multline}
which forces $\ell_1+\ell_2+L = \text{even}$, as required by parity. Finally, setting $s=2$ and comparing with Eq.~\eqref{eq:appB2}, we find 
\begin{multline}
    g^{\text{shear}}_{\ell_1,\ell_2}(L)=\frac{1}{2}\sqrt{\frac{(L+2)!}{(L-2)!}}\sqrt{\frac{(\ell_1+2)!}{(\ell_1-2)!}}\sqrt{\frac{(2L+1)(2\ell_1+1)(2\ell_2+1)}{16\pi}}\begin{pmatrix}
\ell_1 & \ell_2 & L\\
2 & 0 & -2
\end{pmatrix} \frac{1}{2}\left[1+(-1)\right]^{\ell_1+\ell_2+L} \\
\times \frac{1}{\omega_{\ell_1}^2} \frac{C_{\ell_1}^{TT}}{\left(C^{\text{total}}_{\ell_1}\right)^2} \frac{d\ln C_{\ell_1}^{TT}}{d\ln \ell_1} \, .
\end{multline}
This can be made to look closer to its flat-sky counterpart by making use of the recursion relations of the $3j$-symbols to show that
\begin{multline}\label{wig2}
\left[1+(-1)^{(\ell_1+\ell_2+L)}\right]
\begin{pmatrix}
\ell_1 & \ell_2 & L\\
2 & 0 & -2
\end{pmatrix}=\begin{pmatrix}
\ell_1 & \ell_2 & L\\
0 & 0 & 0
\end{pmatrix}\sqrt{\frac{(L-2)!}{(L+2)!}}\sqrt{\frac{(\ell_1-2)!}{(\ell_1+2)!}}\\ \times \left[(\omega^2_L+\omega^2_{\ell_1}-\omega^2_{\ell_2})(\omega^2_L+\omega^2_{\ell_1}-\omega^2_{\ell_2}-2)-2\omega^2_L\omega^2_{\ell_1}\right] \, ,
\end{multline}
where, as discussed in the main text, the term in square brackets on the right accounts for the $\cos (2\theta_{\vec{L},\bell_1})$ weighting in the flat-sky limit. With this simplification, the shear weight function reduces to
\begin{multline}
    g^{\text{shear}}_{\ell_1,\ell_2}(L)=\frac{1}{2}\sqrt{\frac{(2L+1)(2\ell_1+1)(2\ell_2+1)}{16\pi}}\frac{C^{TT}_{\ell_1}}{(C^{\text{total}}_{\ell_1})^2}\frac{d\ln C^{TT}_{\ell_1}}{d\ln \ell_1}\begin{pmatrix}
\ell_1 & \ell_2 & L\\
0 & 0 & 0
\end{pmatrix} \\ \times \omega^2_L
\left[\frac{\left(\omega^2_L+\omega^2_{\ell_1}-\omega^2_{\ell_2}\right)\left(\omega^2_L+\omega^2_{\ell_1}-\omega^2_{\ell_2}-2\right)}{2\omega^2_{\ell_1}\omega^2_{L}}-1\right] \, ,
\end{multline}
given as Eq.~\eqref{eq:gshearcurved} in the main text.

\bibliography{apssamp}
\nocite{*}

\end{document}